\documentclass[12pt]{elsart}
\usepackage[cp1250]{inputenc}
\usepackage[english]{babel}

\usepackage{amsmath}
\usepackage{graphicx}
\usepackage{xcolor}

\newcommand{\Cerenkov}{$\rm \check{C}$erenkov$\,$}

\begin{document}
\begin{frontmatter}
\title{Using \Cerenkov radiation for measuring the refractive index in thick samples by interferometric cathodoluminescence}
\author[USTEM,IFP]{Michael St\"oger-Pollach}\footnote{corresponding author: e-mail: stoeger@ustem.tuwien.ac.at, Tel.: +43 (0)1 58801 45204, Fax: +43 (0)1 58801 9 45204}
\author[USTEM,IFP]{Stefan Löffler}
\author[IFP]{Niklas Maurer}
\author[CEITEC]{Kristýna Bukvi$\rm\check{s}$ová}

\address[USTEM]{University Service Center for Transmission Electron Microscopy (USTEM), Technische Universit\"at Wien, Wiedner Hauptstra\ss e 8-10, 1040 Wien, Austria}
\address[IFP]{Institute of Solid State Physics, Technische Universit\"at Wien, Wiedner Hauptstra\ss e 8-10, 1040 Wien, Austria}
\date{\today}
\address[CEITEC]{Central European Institute of Technology (CEITEC), Brno University of Technology, Purky$\check{n}$ova 123, Brno 612 00, Czech Republic}

\begin{abstract}
Cathodoluminescence (CL) has evolved into a standard analytical technique in (scanning) transmission electron microscopy. CL utilizes light excited due to the interactions between the electron-beam and the sample. In the present study we focus on \Cerenkov radiation. We make use of the fact that the electron transparent specimen acts as a Fabry-Pérot interferometer for coherently emitted radiation. From the wavelength dependent interference pattern of thickness dependent measurements we calculate the refractive index of the studied material. We describe the limits of this approach and compare it with the determination of the refractive index by using valence electron energy loss spectrometry (VEELS).

\end{abstract}

\end{frontmatter}
\section{Introduction}
In recent years cathodoluminescence (CL) in a (scanning) transmission electron microscope (S/TEM) has attracted more and more interest, because its energy resolution is independent from the electron source. The only parameter influencing the energy resolution is given by the analyzing grating. Thus, energy resolutions of $µ$eV are routinely available. Although such high energy resolution is attractive, CL suffers from some drawbacks. These are the limited range of observable energies -- which are usually in the range from infra-red (IR) to soft ultra-violet (UV), which is app. 1~-~4~eV (corresponding to wavelengths in the range of app.~1200~-~300~nm) -- and Fabry-Pérot interference in thin slab-like specimens \cite{yamamoto1996JElecMic,yamamoto2001,stoeger2019um}. Additionally, the small observed energy transfers are strongly delocalized due to the long-range action of the Coulomb force. Thus, the spatial resolution is also limited. When observing incoherent CL, which is the light being emitted after an electron-hole pair recombination, diffusion of electrons and holes contributes further to a decreased spatial resolution. For many applications, a spatial resolution in the range of a few nanometers is sufficient, as can be found in mineralogy (see for example: \cite{silver2015JPCS}) and plasmonics (see for example: \cite{kociak2014a,kociak2014b,horak2018SciRep,horak2019SciRep}).

In the present work, we focus on coherent emission of light inside the S/TEM. This means that the electron beam is directly responsible for the creation of photons and no detour via the process of electron excitation and de-excitation is required \cite{garcia2010RMP}. For this purpose we investigate the light emission of MgO  below the band gap energy. Due to the fact that the refractive index $n$ for visible light is between 1.81 and 1.86 \cite{stephens1952JRNBS}, we make use of the \Cerenkov effect to coherently create photons inside the sample. 200~keV electrons having a speed of $0.695 \cdot c_0$ -- with $c_0$ being the vacuum speed of light -- are fast enough to fulfill the conditions for \Cerenkov light emission \cite{cerenkov1934DAN}, hence we can use this effect as light source. The light is created inside the sample and is partially reflected on the lower and upper sample surface. That way, the sample itself acts as a Fabry-Pérot interferometer. 

On the other hand, if valence electron energy loss spectrometry (VEELS) shall be utilized for the determination of the refractive index \cite{stoeger2008um,stoeger2008micron}, the excitation of \Cerenkov photons is undesired. Any excitation process causes an energy loss and therefore the emission of \Cerenkov photons causes a \Cerenkov loss in the VEELS spectrum. Consequently, for VEELS experiments beam energies of $\leq$~80~keV have to be employed in order to be far below the \Cerenkov limit \cite{horak2015um} of MgO and in order to prevent the VEELS spectrum from being altered due to \Cerenkov losses. 

\section{The TEM specimen as Fabry-Pérot interferometer}

\Cerenkov photons have to be emitted \cite{cerenkov1934DAN}, as soon as the electron traverses the specimen with a velocity $v_e$ faster than the phase velocity of light inside the specimen $c_n = c_0/n(\lambda)$, with $n(\lambda)$ as the wavelength dependent refractive index of the material and $c_0$ the vacuum speed of light. In the present CL study we use MgO and an electron beam energy of 200~keV. This beam energy is high enough to generate \Cerenkov photons. The emission angle of the \Cerenkov photon with respect to the electron trajectory is given by

\begin{equation}
\cos\vartheta_C(\lambda) = \frac{c_0}{v_e n(\lambda)} = \frac{1}{\beta n(\lambda)}.
\label{Cerenkovwinkel}
\end{equation}

\begin{figure}[h!]
\begin{center}
\includegraphics[width=6cm]{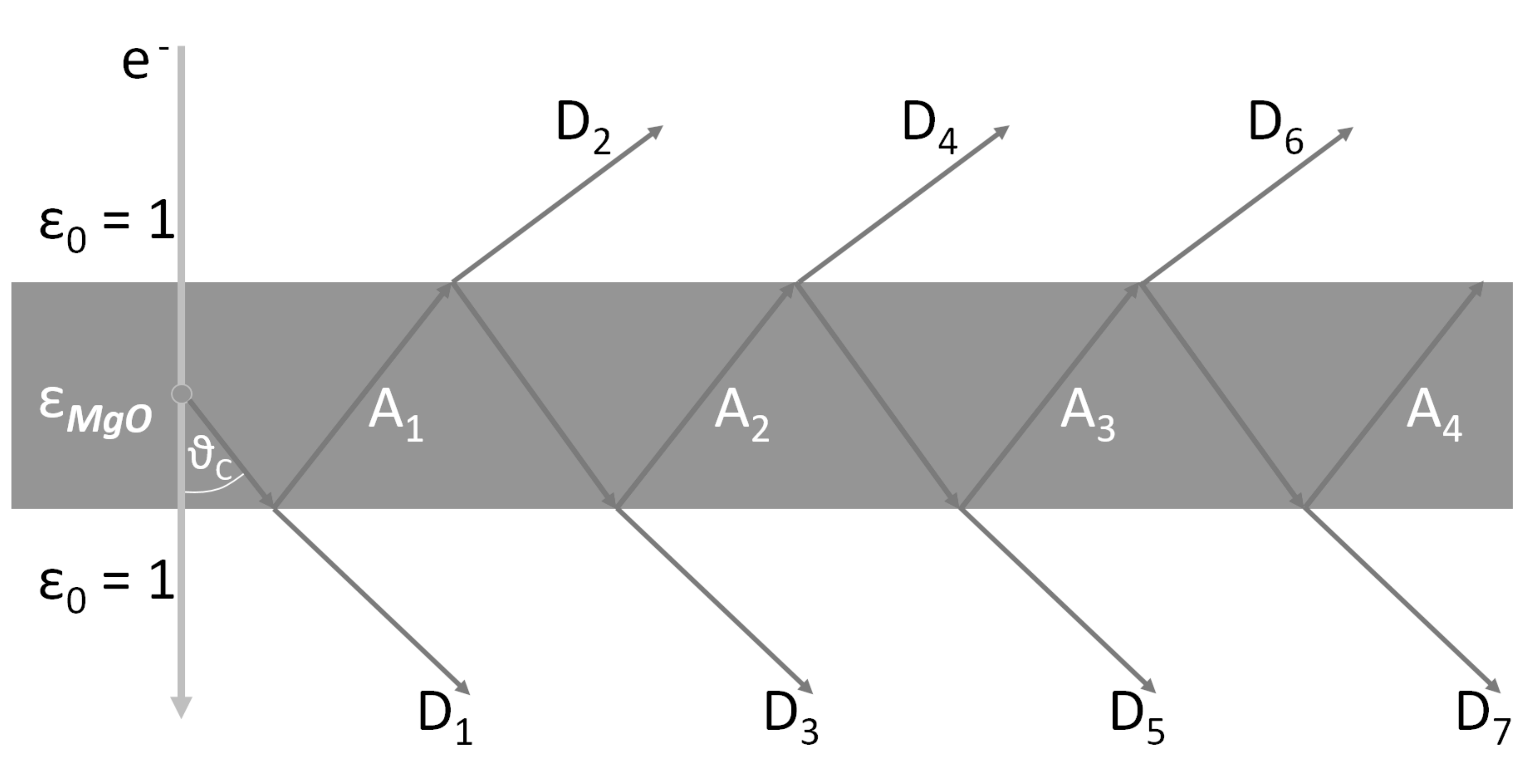}
\end{center}
\caption{Schematic illustration of the TEM specimen acting as a Fabry-Pérot interferometer. The swift electron beam excites \Cerenkov photons which are emitted under the \Cerenkov angle $\vartheta_C$. The partial waves $A_m$ are interfering with each other, $D_{m}$ are emitted.}
\label{FabryPerot}
\end{figure}

Even though the electron beam enters the sample with normal incidence, the created light propagates inside the specimen under a certain angle $\vartheta_C(\lambda)$ (being called $\vartheta_C$ further on). Fig.\ref{FabryPerot} shows a schematic illustration of the experimental geometry. The MgO specimen has a certain dielectric function $\varepsilon_{MgO}$ larger than $\varepsilon_0 = 1$. Therefore the angle of total inner reflection for e.g., $\lambda = 504$~nm is given as

\begin{equation}
\sin \alpha = \frac{\sqrt{\varepsilon_0}}{\sqrt{\varepsilon_{MgO}}} = \frac{1}{1.83} = 0.546. 
\label{Totalreflexionswinkel}
\end{equation}

Consequently, the angle for total inner reflection is 33$^{\circ}$. When following Eq.(\ref{Cerenkovwinkel}), the corresponding \Cerenkov emission angle is 38.16$^{\circ}$ at the same time. Thus, we are facing at least partial total inner reflection and consequently we have to treat the specimen as a Fabry-Pérot interferometer (Fig.\ref{FabryPerot}). The inner reflection is only partially, because of the surface roughness. In some earlier studies the same phenomena were observed, but without making use of it for the determination of the sample's refractive index \cite{stoeger2019um,tizei2013JPhysCondMat}.

The difference in the optical path length $\Delta l$ between the optical rays $A_m$ and $A_{m+1}$ for a specimen of given thickness $d$ is

\begin{equation}
\Delta l = 2d\sqrt{n^2-\sin^2(\vartheta_C)}. 
\label{Wegunterschied}
\end{equation}

Because there is reflection at the optically thinner medium, the phase difference $\Delta \phi$ is

\begin{equation}
\Delta \phi = \frac{2\pi \Delta l}{\lambda} = \frac{4\pi d}{\lambda}\sqrt{n^2-\sin^2(\vartheta_C)}. 
\label{Phasenunterschied}
\end{equation}

Using $\sin^2(\arccos(x)) = 1 - x^2$, substituting $u = 4 + (m\lambda / d)^2$, and using the condition for constructive interference $\Delta \phi = 2m\pi$ -- with $m$ being a positive integer -- $n(\lambda)$ yields ($\beta = v_e/c_0$)

\begin{equation}
n(\lambda) = \pm \frac{1}{2\sqrt{2}}\cdot \sqrt{u\pm \sqrt{u^2-\frac{64}{\beta^2}}}
\label{Kerngleichung}
\end{equation}

The physically meaningful solution is the one having the positive signs. We see in Eq.\ref{Kerngleichung} that an accurate knowledge of $u$ and thus of the sample thickness $d$ at the position of measurement is of utmost importance.

\section{Experimental}

The MgO single crystalline specimen (MaTecK, 99.99\% purity) was prepared by mechanical grinding and a final mechanical lapping procedure, in order not to introduce beam damage caused by further ion milling. Beam damage would be responsible for defect states, thus leading to spurious signals in the incoherent contribution of the CL spectrum. The grinding machine was adjusted to give a wedge angle of 1.3$^\circ$, which was verified by optical measurements.

\begin{figure}[h!]
\begin{center}
\includegraphics[width=6cm]{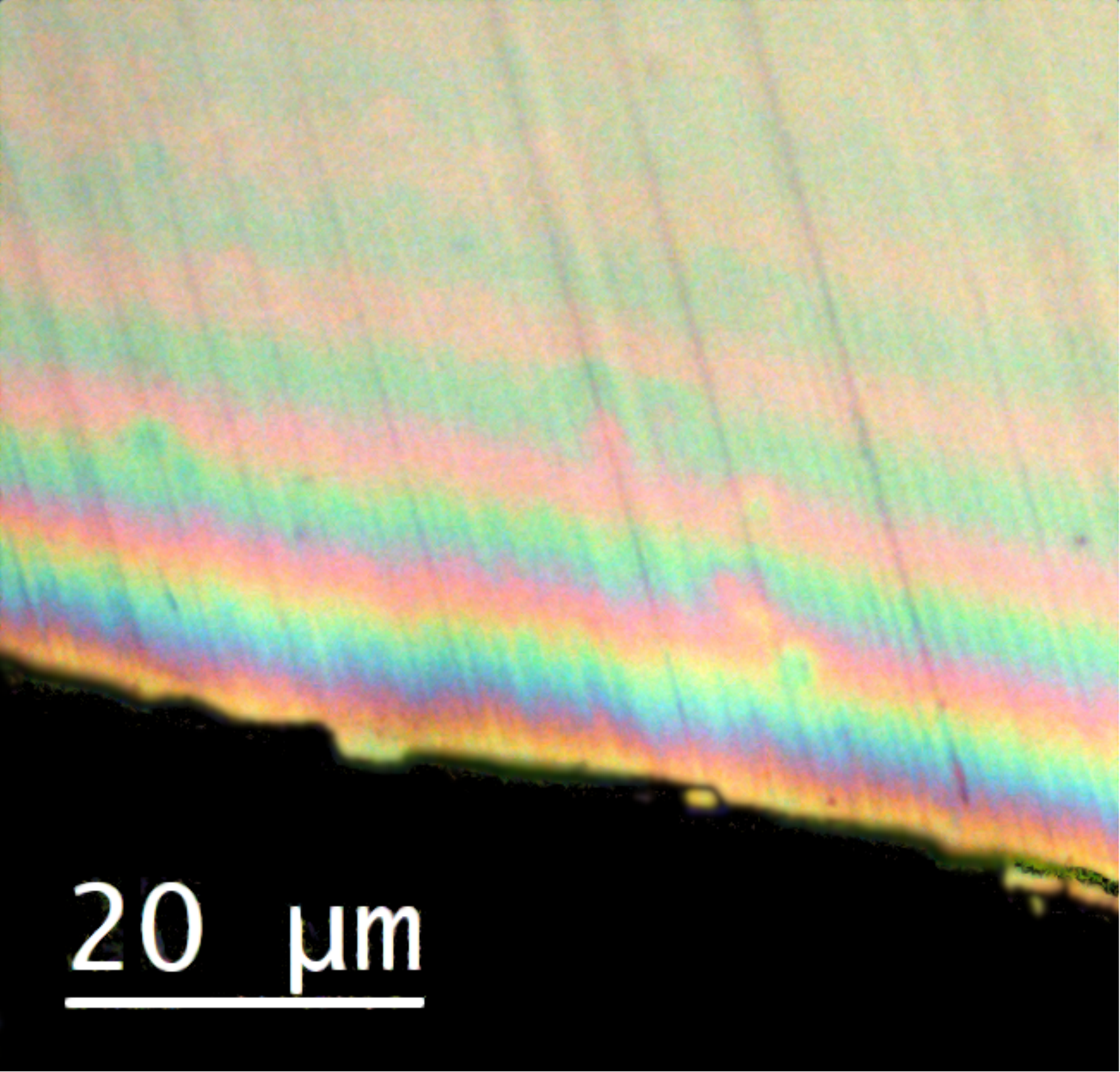}
\end{center}
\caption{Light microscope micrograph of the wedge shaped MgO sample. The wedge angle is 1.3$^\circ$ as adjusted and verified by measuring the distance of the  interference fringes.}
\label{opt}
\end{figure}

The CL study was performed by employing a TECNAI F20 FEG TEM using the scanning mode (STEM) at a beam energy of 200~keV. The high beam energy guarantees the excitation of \Cerenkov radiation acting as light source. A GATAN Vulcan CL detection and analysis system was used.

For the EELS experiments, the beam energy was reduced to 80~keV in order to prevent the excitation of \Cerenkov radiation. \Cerenkov radiation would alter the EELS spectrum due to \Cerenkov losses being present in the band gap of the low loss part in the spectrum \cite{stoeger2006micron}. Albeit \Cerenkov losses can be treated mathematically \cite{stoeger2008micron}, avoiding them is the better solution for an accurate Kramers-Kronig Analysis (KKA) of the valence EELS (VEELS) spectrum \cite{stoeger2015um_a}.

\begin{figure}[h!]
\begin{center}
\includegraphics[width=6cm]{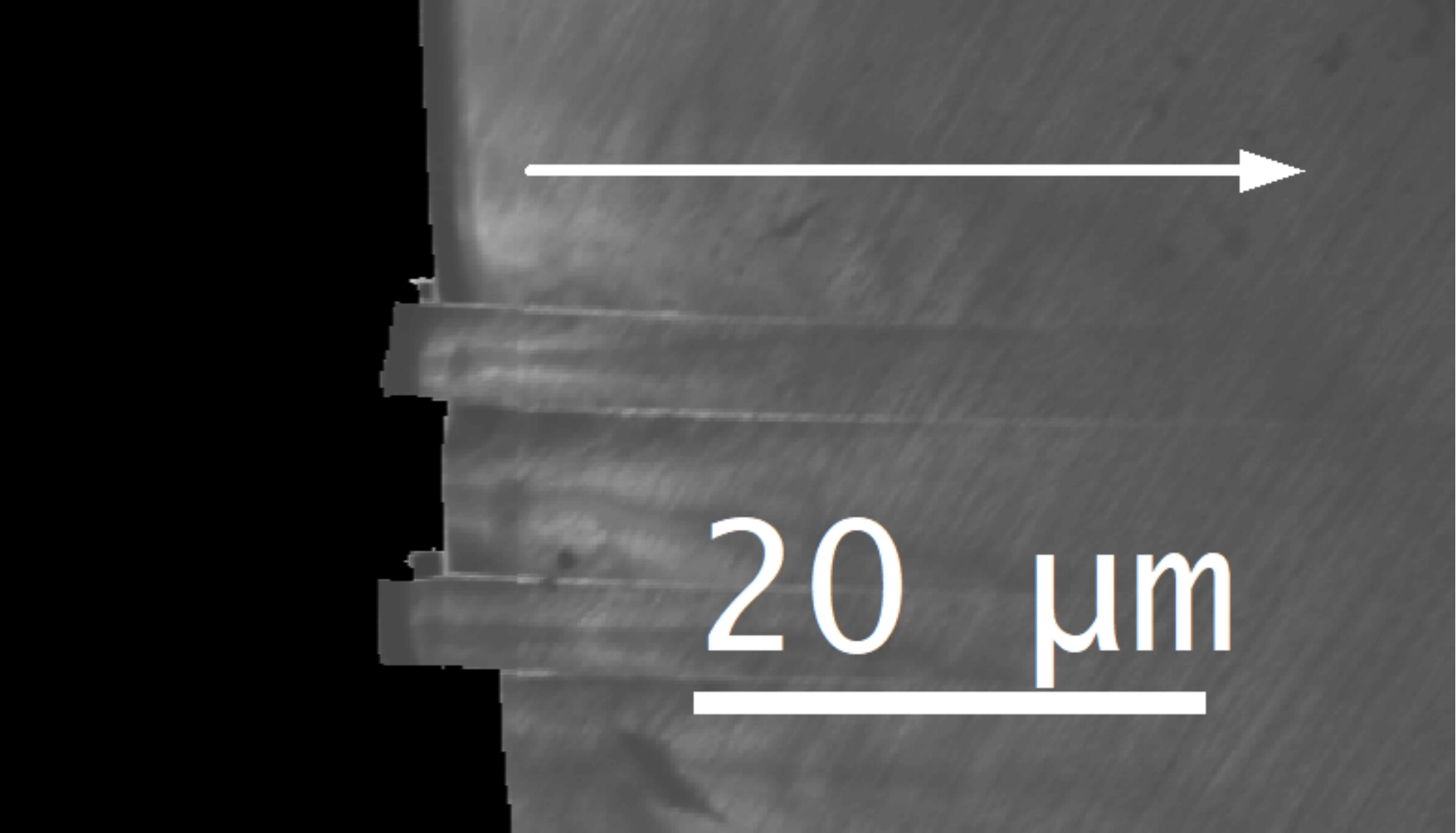}
\end{center}
\caption{High-angle annular dark field image of the observed sample. The white arrow marks the position of the CL line scan acquisition with 300~points of measurement over a length of 30~$\mu$m.}
\label{survey}
\end{figure}

Determining the sample thickness carefully by using EELS at the position of the CL investigations failed because of the large thickness required for the CL experiments. In this case, multiple scattering becomes dominant, thus forming a Landau-background. In such situation any log-ratio method \cite{egerton2012} and multiple scattering deconvolution routines \cite{Schatt84} fails. From the optical investigation in Fig.\ref{opt}, a linear increase in thickness is justified which is further corroborated by the fact that no bending of interference maxima in the CL spectra with respect to sample thickness is observable (see Fig.\ref{CLdata}).

\section{Cathodoluminescence}

For the CL experiments the wedge shaped sample was carefully adjusted into the focal point of the elliptical mirrors of the GATAN Vulcan CL detection and analysis system. The TEM was operated in scanning mode and the beam was scanned across the sample at a length of 30~$\mu$m (Fig. \ref{survey}). Every 100~nm, a CL spectrum was recorded. Due to the wedge shape of the specimen with a slope of 1.3~$^{circ}$, the sample thickness increased by 680~nm from the start-point to the endpoint of the scan. 

\begin{figure}[ht]
\begin{center}
\includegraphics[width=6cm]{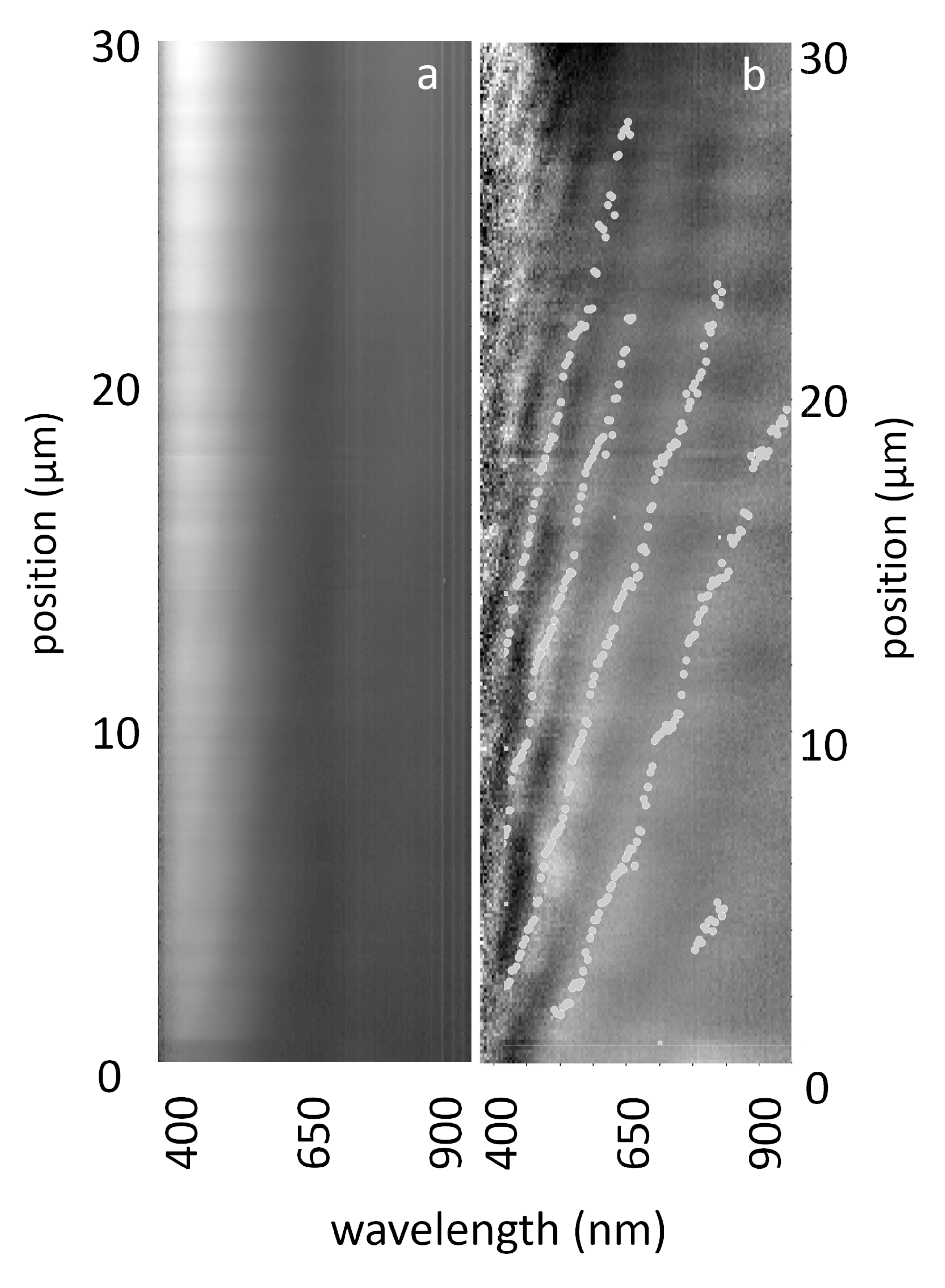}
\end{center}
\caption{(a) As acquired CL spectra with respect to the beam position. (b) Contribution of interference after removal of incoherent CL signals with respect to the beam position. The bright dots mark the measured interference maxima.}
\label{CLdata}
\end{figure}

\begin{figure}[h!]
\begin{center}
\includegraphics[width=6cm]{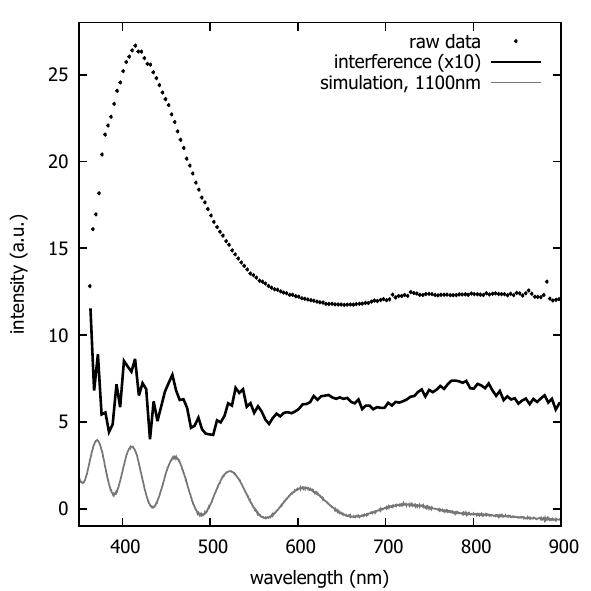}
\end{center}
\caption{Raw CL spectrum (after correction for the system response function) recorded at a sample thickness of 1100~nm and its contribution from interference after removal of the incoherent spectral fraction. The simulation confirms the interpretation of the observed fringes.}
\label{CLspectra}
\end{figure}

Figure \ref{CLdata}a shows the acquired spectrum image containing all CL spectra with respect to the beam position on the sample. There are two prominent features already visible: (i) faint interference fringes and (ii) intensities at 704~nm, 725~nm, 860~nm, 880~nm, 904~nm, and 922~nm which are constant in energy at all thicknesses. The former are caused by interference, the latter are caused by impurity levels inside the band gap. When removing the incoherent contribution, which is of equal shape in all spectra, only the spectral fraction stemming from interference remains (Fig.\ref{CLdata}b). As example Fig.\ref{CLspectra} shows a single measurement at the 1100~nm thick MgO position, its contribution stemming from interference after removal of all incoherent parts, and the corresponding simulation based on Yamamoto's theory \cite{yamamoto1996JElecMic,yamamoto2001,stoeger2019um} using optical data \cite{stephens1952JRNBS}. The CL spectrum was corrected for the system response function. This system response includes the reflectivity of the mirrors, the absorption of the light guides, the correction for the 500~nm blazed grating, and the wavelength dependent detection quantum efficiency of the CCD detector.

\begin{figure}[h!]
\begin{center}
\includegraphics[width=6cm]{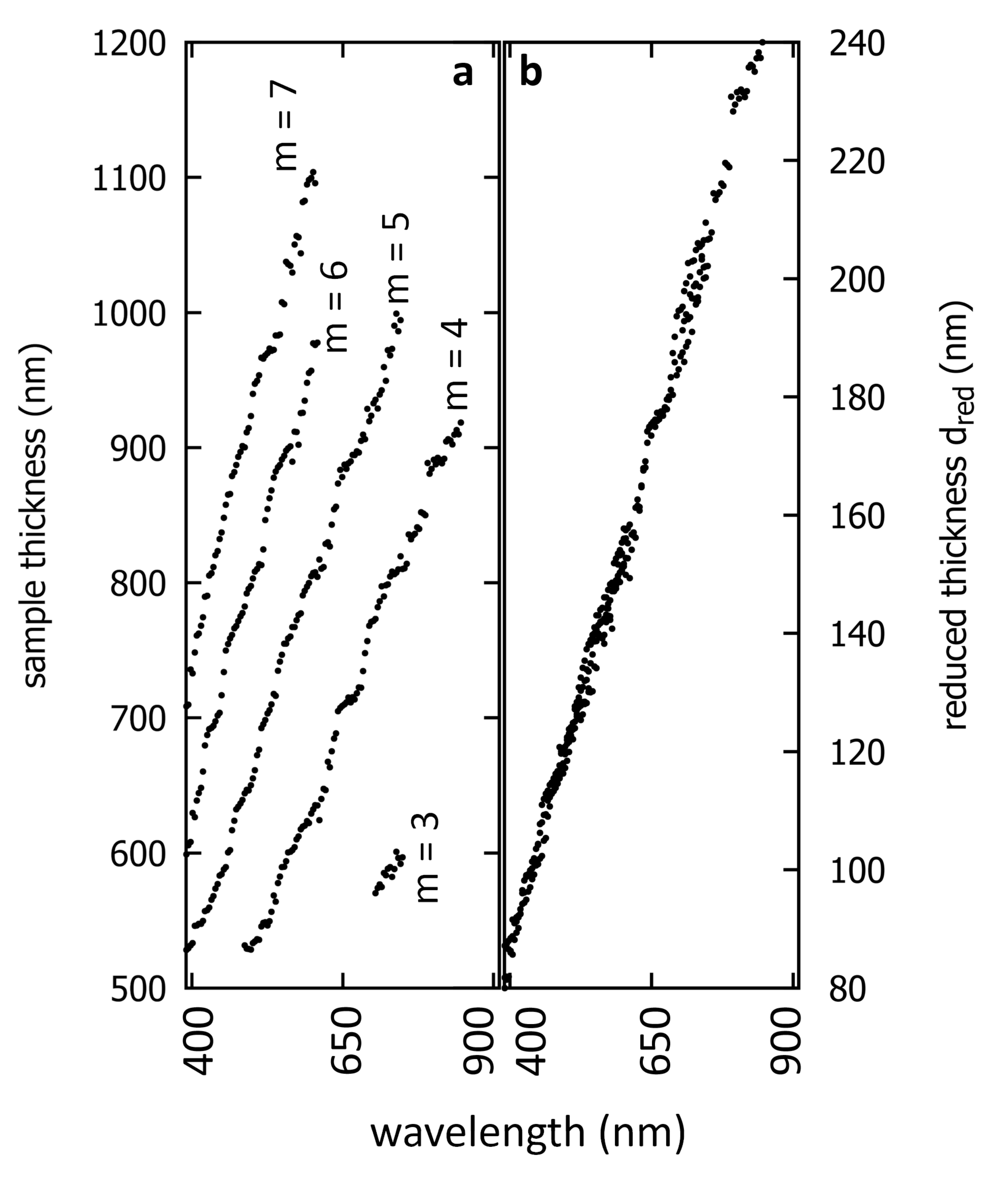}
\end{center}
\caption{(a) Identified maxima for given $m$ with respect to the sample thickness. (b) The reduced sample thickness $d_{red}=d/m$ which is comparable to the first interference maximum $m = 1$.}
\label{max}
\end{figure}

For any interference experiment, the total thickness of the sample is of utmost importance. Due to the fact that thick samples are needed for this kind of experiments, thickness determination by means of EELS fails. Therefore we make use of the interference pattern in CL and the knowledge about the sample geometry. In the present case the sample was prepared having a wedge angle of 1.3$^\circ$ being confirmed with optical methods. Additionally, $m$ has to be a positive integer. Using these two boundary conditions we can make an educated guess for the sample thickness at the starting-point of the line scan. In the present experiment, we estimated 580~nm sample thickness. Subsequently we varied the sample thickness during the calculation of the reduced thickness $d_{red}=d/m$ for all interference maxima (Fig.\ref{max}). At the correct thickness, all interference maxima should be on a single curve, independently of their interference order $m$ (Fig. \ref{max}b). Finally, the minimum median of the standard deviation of the reduced thicknesses with respect to the single curve is giving the accurate sample thickness at the starting-point. In the present experiment the thickness at the starting-point is found to be 560~nm.

\begin{figure}[h!]
\begin{center}
\includegraphics[width=6cm]{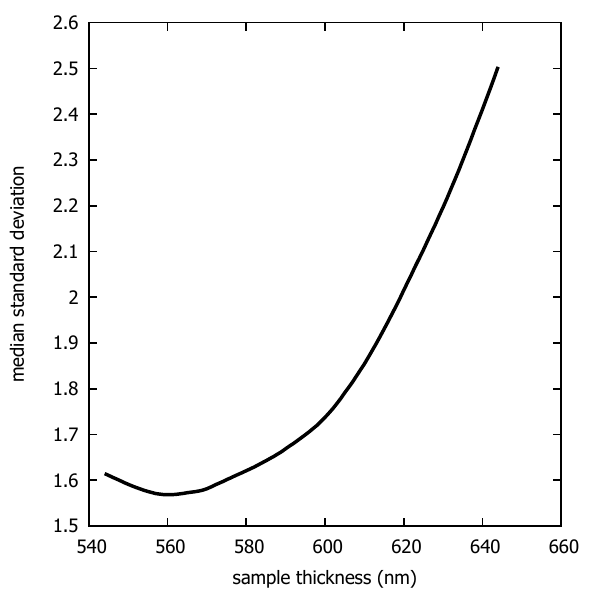}
\end{center}
\caption{Median standard deviation of $d_{red}=d/m$ for all wavelengths for various estimated sample thicknesses.}
\label{statistik}
\end{figure}

\begin{figure}[h!]
\begin{center}
\includegraphics[width=6cm]{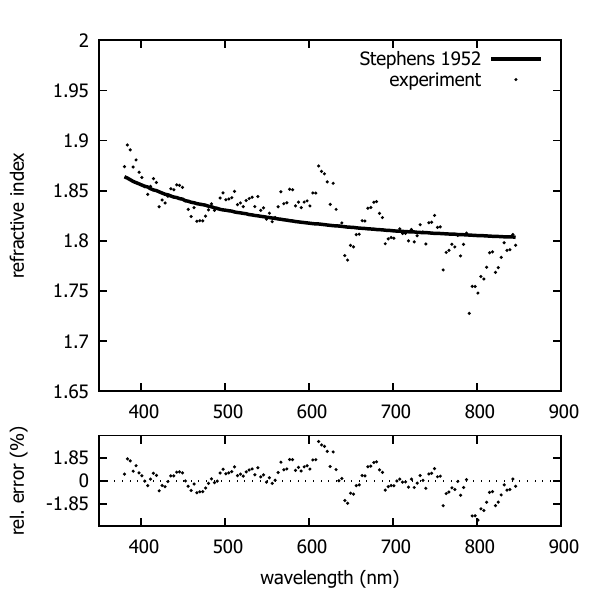}
\end{center}
\caption{Experimentally determined refractive index in comparison with optical reference data from \cite{stephens1952JRNBS} and the error of the method with respect to the optical reference data.}
\label{result}
\end{figure}

Inserting the results shown in Fig.\ref{max}b into Eq.\ref{Kerngleichung} and taking $u$ as being $u = 4 + (\lambda/d_{red})^2$ one can calculate the refractive index for the probed range of wavelengths. Figure \ref{result} gives the refractive index $n$ in comparison with data from literature \cite{stephens1952JRNBS}. In the lower panel, the relative error is given. The error is mainly due to the scratches from the mechanical preparation already visible in Fig. \ref{opt}.

\section{Valence EELS}

\begin{figure}[h!]
\begin{center}
\includegraphics[width=6cm]{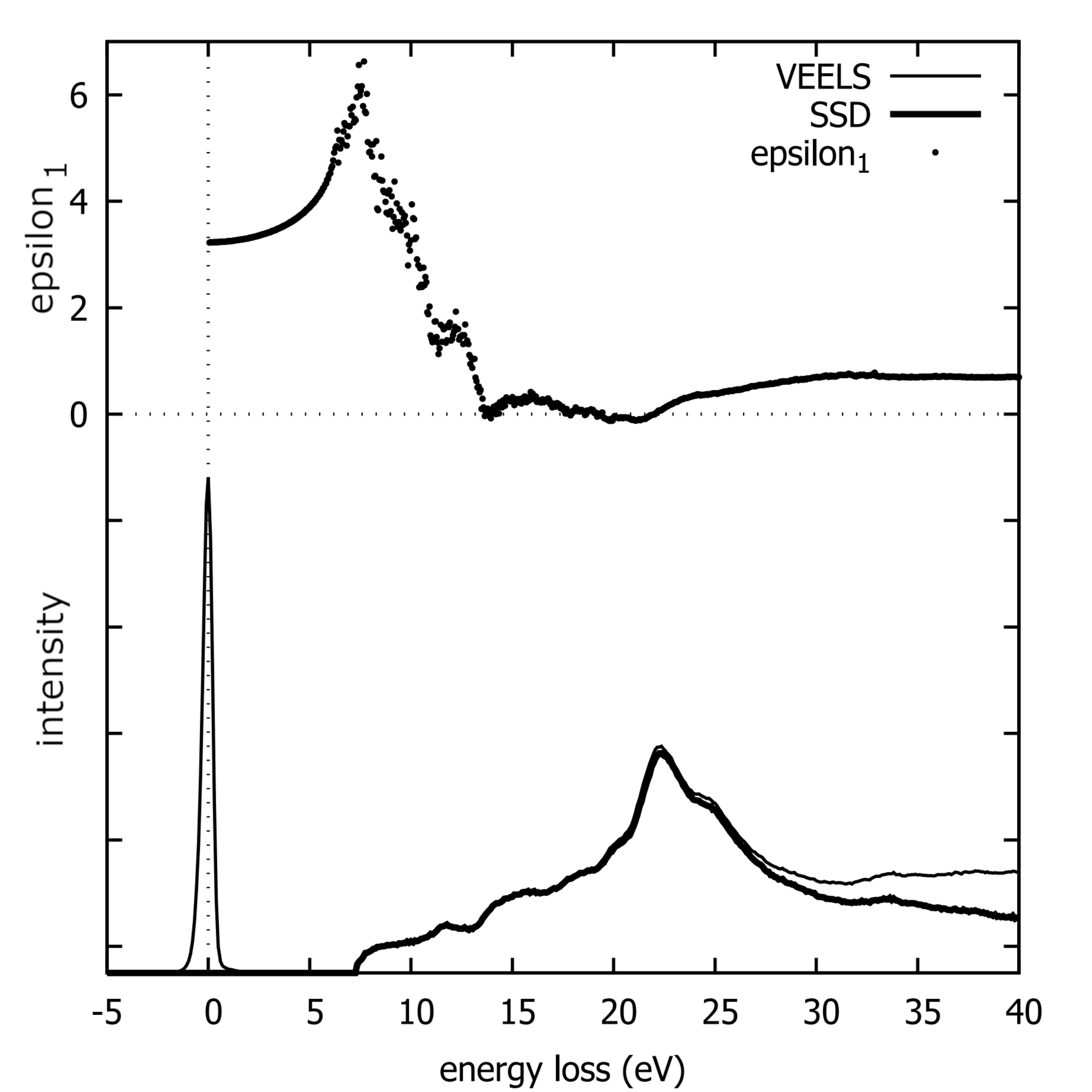}
\end{center}
\caption{80~keV VEELS spectrum of 82~nm thick MgO and its SSD. By means of KKA, the dielectric function was determined -- here only the real part $\varepsilon_1$ is plotted.}
\label{VEELSresult}
\end{figure}

The VEELS experiments were performed at 80~keV at a TECNAI G20 LaB$_6$ in order to suppress any \Cerenkov losses. The spectrometer resolution was 0.53~eV as full width at half maximum of the zero loss peak (ZLP) at the fully saturated LaB$_6$ filament. For data analysis, an 82~nm thin areas of the ample was selected. The thickness was determined via the log-ratio method \cite{egerton2012}. Data analysis included multiple scattering deconvolution based on the Fourier-log method utilizing a pre-measured vacuum ZLP. That way, the single scattering distribution (SSD) was retrieved. The band gap is the onset of the inelastic signal and is identified to be 7.3~eV \cite{heo2015AIPAdv}. Subsequent Kramers-Kronig Analysis (KKA) was performed using the sample thickness for normalization \cite{stoeger2008micron}. The resulting complex dielectric function $\varepsilon_1 + i\varepsilon_2$ was further used for calculating the refractive index $n$. In Fig.\ref{VEELSresult} the raw experimental spectrum, its SSD and the real part of $\varepsilon$ are shown.

Due to the fact that by using CL, we are limited to visible light and thus to energies smaller than the band gap energy of MgO, we have a closer look to the results of KKA only in the range from 1.2~-~4.0~eV. Within the band gap, the absorption coefficient $\kappa$ and thus the imaginary part of $\varepsilon$ are zero. Therefore the square root of $\varepsilon_1$ is equal to the refractive index $n$.

\begin{figure}[h!]
\begin{center}
\includegraphics[width=6cm]{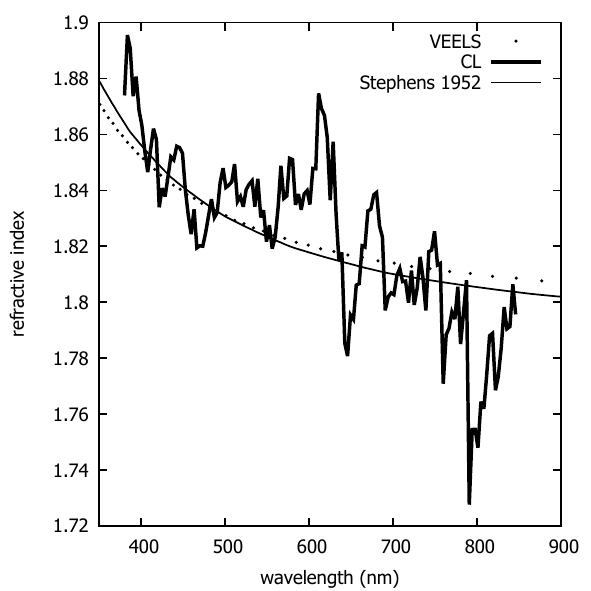}
\end{center}
\caption{Refractive index calculated from VEELS and CL data compared to the optical reference \cite{stephens1952JRNBS}.}
\label{VEELScomparison}
\end{figure}

In this situation the advantage of VEELS is, that within the band gap $\varepsilon_2 = 2n\kappa = 0$. Hence $\varepsilon_1$ is a smooth function. Whereas in CL we are restricted to the quality and smoothness of the sample surface, in VEELS such properties do not dominate the resulting refractive index. Consequently, VEELS is preferable to interferometric CL as long as thin samples are be investigated. But when investigating optical resonators and light guiding particles of a certain thickness not suitable for EELS, interferometric CL gives the possibility for the determination of the optical properties.

\section{Discussion}

In general \Cerenkov radiation is seen as a disruptive signal in CL as well as in VEELS. In the present study we make use of the fact, that \Cerenkov radiation is reflected inside the specimen, thus generating a wavelength and sample-thickness dependent interference spectrum. The sample itself acts as a Fabry-Pérot interferometer, hence the refractive index can be determined with high energy resolution. The signal-to-noise ratio of the CL spectra determines the accuracy of this method. Anyhow, VEELS at low beam energies is preferable in the case of MgO.

\section*{Acknowledgements}

This research was in part supported by the SINNCE project of the European Union's Horizon 2020 programme under the grant agreement No. 810626.

\bibliographystyle{elsart-num}

\end{document}